\documentstyle[psfig]{caosp}
\begin{document}
%\pubyear{1998}
%\volume{27}
%\firstpage{365}
\htitle{Atmospheric abundances of CP SB2 star components of equal masses.
       II. 66 Eridani}
\hauthor{A.V.Yushchenko{it et al.}}
    \thesaurus{ }

\title{Atmospheric abundances of CP SB2 star components of equal masses.
       II. 66 Eridani    }
\author{A.V. Yushchenko \inst{1} \and
        V.F. Gopka      \inst{1} \and
        V.L. Khokhlova  \inst{2} \and
        F.A. Musaev     \inst{3} \and
        I.F. Bikmaev    \inst{4} }
   \offprints {A.V. Yushchenko, $e-mail$: yua@lens.tenet.odessa.ua }
   \institute{    Odessa Astronomical Observatory, Odessa, Ukraine
             \and Institute of Astronomy,  Moscow, Russia
             \and Special Astrophysical Observatory, Zelenchuck, Russia
             \and Kazan State University, Kazan, Russia }

   \date{  }
   \maketitle{}
   \begin{abstract}
    We report  the  results  of abundance  determination for the
 components of the SB2 star 66 Eri ($M_{\rm A}/M_{\rm B}$=0.97) from high
 resolution CCD
 echelle spectra  with S/N{$\ge$}100  taken with the 1-m telescope of Special
 Astrophysical Observatory (Zelenchuck, Russia).
The atmospheric parameters of the components were determined  using all
 available photometric, spectrophotometric and spectral data.
 The abundances of~ 27 elements were found.
   The abundances of components  are  different.
The B component, previously  classified as an Hg-Mn star, does not show
anomalies typical of this group such as deficit of He,  Al and excess
of P, Ga  but shows  overabundances  of heavy elements
which amount  up to 4-5 dex.
The A component also shows moderate Mn and Ba excess. Lines of other heavy
elements were not detected. Estimates of upper limits to their abundances do
not permit to exclude completely the presence of fainter anomalies
in the A component either.
\end{abstract}

 This work is part of our programme to investigate atmospheric abundances
of SB2 system components of equal masses.
The spectroscopic orbit of 66 Eri (HD 32964)  was published by Young (1976).
The mass ratio of the B9V+B9V components, $M_{\rm A}/M_{\rm B}=0.97$ is
the closest to unity among the systems which are investigated in this
programme. The orbital period of this binary is 5.522731$^d$.
 66~Eri is designated as a mercury star in Yale Bright Star Catologue,
 but as a variable star (EN~Eri) of the $\alpha^2$~CVn type in the Catalogue
of Variable stars with a lightcurve amplitude of 0.005$^{m}$ and a photometric
period of 7.86$^{d}$ (Schneider, 1987).

We analyzed two spectra of 66 Eri obtained with the
\'echelle spectrograph of the 1-m telescope (Musaev, 1996)
of the Special Astrophysical Observatory at Zelenchuk (Russia)
in the wavelength region 4385-6695 \AA,~
with a signal-to-noise ratio S/N$\geq $100 and a spectral resolution of 36000.
The reduction of the spectra was done with the DECH code (Galazutdinov, 1992) 
and URAN package written in Odessa observatory.

  As the components of 66 Eri are very similar,
 we can use photometric calibrations for single stars to
 determine mean characteristics of the system.
      We made flux calculations with an abundance pattern typical of CP stars
  and Kurucz' (1993) line list ($>$31 millions lines).
  The result obtained from
  photometric calibrations and from comparison of observed
  and calculated fluxes
  gave us $T_{\rm eff}=11000$~K, $\log g=4.25$ as mean atmospheric parameters.

  We tried to find $T_{\rm eff}$, $v_{\rm turb}$ and correction factor of
  equivalent widths $WCF$ for each component
  using unblended lines of Fe\,{\sc ii} in both spectra.
   We took the grid of values of $T_{\rm eff}$, $v_{\rm turb}$, $WCF$,
fixed $\log g$
  and
  determined the Fe abundance of each component for each point of
  the grid and then selected parameters,
  which met three conditions simultaneously:
   zero correlation between equivalent widths
       and  calculated abundances;
   zero correlation between energies of the lower levels and
        calculated abundances;
   minimal r.m.s. error for iron abundance.
  This has permitted  us to select the parameters:
 $T_{\rm effA}=11100$~K, $v_{\rm turbA}=0.9$~km\,s$^{-1}$, $WCF_{\rm A}=1.95$;
 $T_{\rm effB}=10900$~K, $v_{\rm turbB}=0.7$~km\,s$^{-1}$, $WCF_{\rm B}=2.05$.

  The projected rotational velocity of the components was found to be
  $v\sin i=17$~km\,s$^{-1}$.
  The parallax of 66 Eri, measured
  by the Hipparcos satellite, is $\pi =11.65\pm 0.73$ mas
  (Perryman et al., 1997).
 Combining this value with the visual magnitude and flux ratio of the system,
 and with bolometric corrections and effective
  temperatures of the  components, we found the radii of the components
  $R_{\rm A}=1.75$, $R_{\rm B}=1.86$, and
the rotational periods: $P_{\rm A}=5.21\sin i$ and $P_{\rm B}=5.53\sin i$ days.
These values are close  to the orbital period $P=5.5227^d$, so the rotation
is synchronized if $\sin i$ is close to 1.

\begin{table*}
\begin{center}
\footnotesize {
\caption{Atmospheric abundances for both components of 66 Eri}
%\centerline {
\begin{tabular}{| l |  rrr | rrr | rr | }
\hline
   Ident.   & $n_A$ & $logN_A$ & $\sigma_A$ & $n_B$ & $logN_B$ &
                            $\sigma_B$          &
                            $A-\odot$  & $B-\odot$ \\
\hline
 He I  &  6 &10.98& .09& 5 &10.98& .03& -.01 &  -.01 \\
 C  I  &  1 & 8.40&    & 1 & 8.37&    & -.16 &  -.19 \\
 O  I  &  9 & 8.79& .03& 7 & 8.68& .01& -.14 &  -.25 \\
 Ne I  &  1 & 8.15&    &   &     &    &  .06 &       \\
 Mg I  &  6 & 7.49& .07& 5 & 7.15& .07& -.09 &  -.43 \\
 Mg II &  6 & 7.49& .17& 5 & 7.28& .13& -.09 &  -.30 \\
 Al II &  4 & 6.64& .09& 2 & 6.32& .11&  .17 &  -.15 \\
 Si II &  7 & 6.97& .19& 5 & 6.87& .07& -.58 &  -.68 \\
 S  II &  2 & 7.34& .13& 1 & 7.26&    &  .13 &   .05 \\
 Ca II &  2 & 6.62& .15& 2 & 6.36& .12&  .26 &       \\
 Sc II &  1 & 3.35&    & 1 & 3.14&    &  .25 &   .04 \\
 Ti II & 22 & 5.24& .21&27 & 5.90& .16&  .25 &   .91 \\
 Cr I  &    &     &    & 1 & 6.09&    &      &   .41 \\
 Cr II & 20 & 5.92& .18&18 & 6.42& .16&  .25 &   .75 \\
 Mn II &  2 & 6.00& .18& 3 & 6.39& .20&  .61 &  1.01 \\
 Fe I  &  7 & 7.88& .20& 8 & 7.71& .30&  .24 &   .07 \\
 Fe II & 62 & 7.72& .13&60 & 7.66& .10&  .08 &   .02 \\
 Ni I  &    &     &    & 1 & 6.62&    &      &   .37 \\
 Zn I  &    &     &    & 1 & 5.89&    &      &  1.29 \\
 Y  II &    &     &    &14 & 5.21& .31&      &  2.97 \\
 Zr II &    &     &    & 3 & 4.19& .17&      &  1.59 \\
 Ba II &  4 & 3.41& .06& 6 & 3.82& .13& 1.28 &  1.69 \\
 La II &    &     &    & 2 & 4.05&    &      &  2.83 \\
 Ce II &    &     &    & 1 & 3.95&    &      &  2.40 \\
 Yb II &    &     &    & 4 & 4.32& .15&      &  3.24 \\
 Hf II &    &     &    & 1 & 4.37&    &      &  3.49 \\
 W  II &    &     &    & 4 & 4.34& .24&      &  3.23 \\
 Pt I  &    &     &    & 5 & 6.75& .31&      &  4.95 \\
 Au I  &    &     &    & 1 & 6.66&    &      &  5.65 \\
 Hg I  &    &     &    & 2 & 5.88& .10&      &  4.79 \\
 Hg II &    &     &    & 1 & 6.40&    &      &  5.31 \\
\hline
\end{tabular} %  }
  }
\end{center}
\end{table*}

  For the identification of lines we used
 Tsymbal's (1995) spectrum synthesis programme and
 Kurucz' (1993) line data from CD-ROMs 18, 23.

 We made several iterations with different
 abundance patterns to achieve the best fit of the computed spectrum to
 the observed one in the whole spectral region.
 Only lines free of blending by lines of the other component were taken
 into account.
 Abundances of the elements with Z$\ge$16, except barium, were
 obtained with the technique of equivalent widths using the WIDTH9 code.
 Abundances of He, C, N, Ne, Mg, Al, Si, Ba were determined with the
 method of spectrum synthesis using Kurucz' (1993) SYNTHE programme.
 The abundance of barium was obtained with the inclusion of hyperfine
 structure according to Francois (1996).
 Results  are given in Table 1, where $n$ is the full number of measured
 lines of an element in two spectra, $logN$ is the abundance of an element
 in the scale $logN(H)=12.0$, $\sigma$ is the mean square error of one
 measurement,
 $A,B-\odot$ is the abundance relative to the Sun.

 No correlation was found between the elemental
 abundance and the Land\'e factor for lines of Fe, Ti, Mn and Cr.
 We attempted to identify lines of P, Ga and Xe,
 but no detecteble line was found in the spectra.
To our knowledge 66 Eri is the first SB2 system studied so far which contains
chemically peculiar, but non-HgMn type components.

\acknowledgements
 Part of the work was supported by Russian GNTP ``Astronomy''. The authors
 express their gratitude to Dr. G. Galazutdinov for his help in observations
 and for a new version of his DECH code.


\begin{thebibliography}{}
\article    {Francois, P.}{1996}{\aaa}{313}{229}
\bibitem{}  Galazutdinov, G.: 1992,
            {\it Preprint of Special Astrophys. Obs. RAS}, {\bf 92}
\bibitem{}  Kurucz, R.L.: 1993, CD-ROMs 1-23, Smithsonian Astrophys. Obs.
\article    {Musaev, F.A.}{1996}{\sal}{22}{795}
\bibitem{}  Perryman, V.A.C. et al.: 1997, The Hipparcos and Tycho Catalogues,
            European Space Agency, CD-ROMs 1-8
\bibitem{} Schneider, H.: 1987, {\it Hvar. obs. Bull.} {\bf  11}, No. 1, P. 29
\article    {Tsymbal, V.V.} {1994}{Odessa Astron. Publ} {7} {144}
\article    {Young, A.}{1976}{\pasp}{88}{275}
\end{thebibliography}
\end {document}